\documentclass[aps,prl,showpacs,twocolumn,preprintnumbers]{revtex4}
\usepackage{amssymb}
\usepackage{times}
\usepackage{amsmath}
\usepackage{graphicx}

\begin{document}

\title{Emergence of coexisting coherence and incoherence by an external forcing}

\author{V. K. Chandrasekar$^{1}$}
\author{R. Suresh$^{1}$}
\author{D. V. Senthilkumar$^{1}$}
\email{skumarusnld@gmail.com}
\author{M. Lakshmanan$^{2}$}

\affiliation{
$^{1}$Centre for Nonlinear Science and Engineering, School of Electrical and Electronics Engineering, SASTRA University, Thanjavur  613 401, India\\
$^{2}$Centre for Nonlinear Dynamics, School of Physics, Bharathidasan University, Tiruchirappalli  620 024, India}

\date{\today}

\begin{abstract}

A common external forcing can cause a saddle-node bifurcation in an ensemble of identical Duffing oscillators by
breaking the symmetry of the individual bistable (double-well) unit. The strength of
the forcing determines the separation between the saddle and  node, which in turn
dictates different dynamical transitions depending on the
distribution of the initial states of the oscillators. In particular, chimera-like states
appear in the vicinity of the saddle-node bifurcation for which theoretical
explanation is provided from the stability of slow-scale dynamics of the 
original system of equations. Further, as a consequence, it is shown that 
even a linear nearest neighbor coupling can lead to the manifestation of the chimera states
in an ensemble of identical Duffing oscillators in the presence of the common external forcing.

\end{abstract}

 \maketitle
\newpage
\bigskip

Ensembles of coupled oscillators are veritable black boxes which have been
widely employed to understand  a plethora  of collective/co-operative dynamics 
observed in diverse natural systems~\cite{apmr2001,atw1980,mldvsk2010}.
Identification of simultaneous existence of a group of oscillators in harmony 
while the rest in dissonance in an ensemble of coupled identical oscillators  has
resulted in the notion of ``chimera" states~\cite{ykdb2002}. Since the first 
identification of the coexistence of coherence and incoherence by Kuramoto and  Battogtokh~\cite{ykhn1996},  many exciting
developments have been made, including a number of experimental demonstrations~\cite{mrtsn2012,eamst2013,llbp2013}.
A flurry of recent investigations revealed nonlocal coupling as a basic criterion for
the onset of chimera in phase reduced models under weak coupling limit~\cite{ykdb2002,ykhn1996},
and also in systems far from the weak coupling limit~\cite{ioym2011}.
Very recently, the above limitation on the existing criterion for the onset of 
chimera has been liberalized to include 
an ensemble of globally coupled oscillators~\cite{gsas2014,ayap2014,vkcrg2014}. 
Nevertheless, the mechanism for the birth of chimera states in an 
ensemble of identical oscillators has been shown as the emergence of multistability
in both nonlocal and global coupling configurations~\cite{ykdb2002,ykhn1996,ioym2011,gsas2014,ayap2014,vkcrg2014}.
However, it is not clear whether the converse is true, that is whether chimera 
can occur in a given multistable system either with nonlocal or with global coupling
or even with much  simpler couplings such as linear nearest neighbor coupling.
A general criterion for the onset of chimera states in such dynamical
systems is yet to be unraveled. 

A periodically driven oscillator is the main and historically first studied model
in the classical theory of synchronization, where a triode generator
is synchronized by a weak external periodic signal~\cite{eva1922}. In reality,
there exists a variety of instances where the onset of such a collective behavior is
invoked by an external force~\cite{apmr2001,atw1980,mldvsk2010}. Examples include,
phase locking of electrically decoupled spin torque nano oscillators (STNOs)
by an applied external microwave magnetic field~\cite{skmrp2005,bsvkc2013},
polariton condensates in semiconductor microcavities that interact with
the reservoir~\cite{mw2008}, frequency controlled devices, 
synchronization of micromechanical oscillators using light~\cite{mzgsw2012}, etc. 
Here, we first consider an ensemble of identical bistable systems driven by a common 
external force without establishing any explicit coupling between the systems and
investigate the underlying dynamics of the ensemble
(or equivalently, that of an identical system for an ensemble of different initial conditions) as
a function of the strength of the forcing. Then we investigate the effect of coupling
leading to the manifestation of chimera states.

The collective steady states  resulting from the basin of attraction of the ensemble 
of identical oscillators (or an ensemble of initial conditions)
provides a clue about the basic criterion for the onset of chimeras in the 
ensemble of  such systems with coupling.  In particular, we find that if 
the basin of attraction of a bistable (multistable) system  leads
to coexisting attractors of distinctly different nature under a common forcing, 
the ensemble of such  systems will immediately lead to the 
onset of chimera near the saddle node bifurcation of the ensemble under 
coupling. 
For appropriate forcing amplitude, the basin of attraction of the bistable system
has essentially two domains of attraction.   Initial conditions from one of 
the domains (coherent domain) lead to the same (chronotaxic~\cite{yfsptc2013}) 
attractor while that of the other domain (incoherent domain) lead to 
quasiperiodic/chaotic attractors. Thus the collective steady 
state dynamics, whose initial conditions are uniformly distributed in the two domains, 
emerging from the basin of attraction resembles the chimera states under the common external forcing.  
In the following, we will show  as 
a general criterion that if the basin of attraction of a given dynamical system
displays coexisting coherent and incoherent domains of attraction under 
the common external forcing, then even a linear nearest neighbor coupling
can lead to the manifestation of chimera states in the ensemble of identical 
systems even for a low coupling strength in the same parameter space. 
This approach will be extremely
useful, and provides a cost-effective way in laboratory experiments to confirm
the emergence of chimeras with just a single oscillator before experimenting with an ensemble
of oscillators. In addition, the ensemble of identical oscillators driven by
a common forcing opens up the possibility of analytical treatment which we will
carry out in the following.

In this Rapid Communication, we consider a bistable system with double-well potential as an individual 
oscillator exhibiting reflection symmetry which  undergoes a pitch-fork bifurcation~\cite{sykyk2000}
as a function of the system parameter.
The basin of attraction of the ensemble of  such identical oscillators leads to two 
cluster states corresponding to the bistable attractors with reflection symmetry. 
The common external force then leads to a saddle-node bifurcation of the
ensemble by breaking the reflection symmetry of the individual oscillators. 
For suitable values of the amplitude of the external force near the saddle-node 
bifurcation, we find the co-existence of coherent and incoherent 
steady states among the oscillator ensemble mimicking chimera states. 
Using  a slow-scale 
approximation and the method of direct separation~\cite{af2006}, we separate
the slow-scale dynamics from the original system of equations.  The stability
of the steady state solutions of the slow-scale dynamics provides  an appropriate 
theoretical explanation for the observed  dynamical transitions in the simulation.

\begin{figure}[tbp]
	\centering
	\includegraphics[width=1.0\columnwidth]{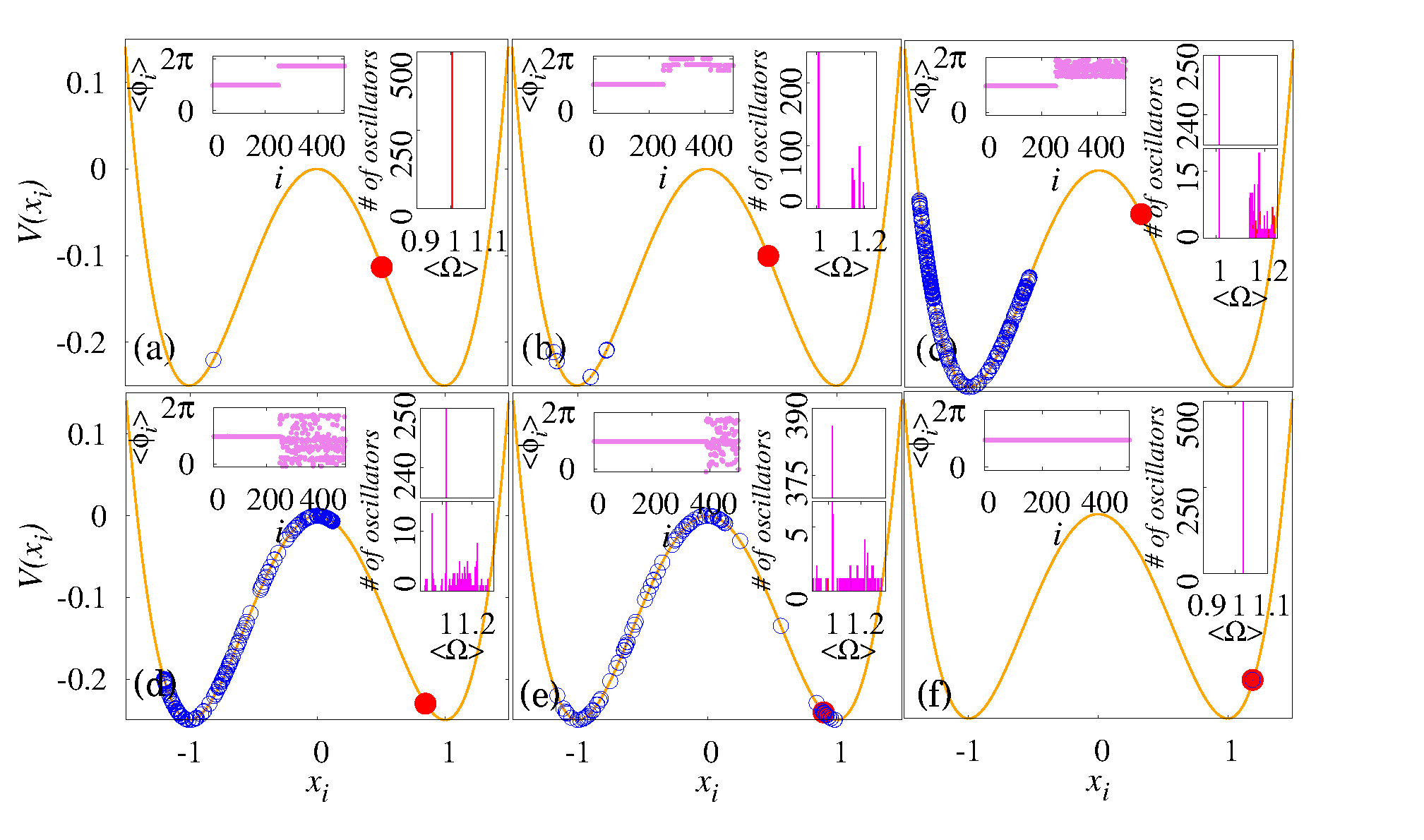}
	\caption{(Color online) Distribution of the ensemble ($N=500$) of Duffing oscillators in a double-well potential along with 
		their average phase and average frequency distribution in the insets. (a) $f_e=0$, (b) $f_e=0.088$, (c) $f_e=0.135$, 
		(d) $f_e=0.15$, (e) $f_e=0.16$ and (f) $f_e=0.19$. See text for explanations.} 
	\label{fig1}
\end{figure}
In general, an oscillator with a double-well potential of the
form $V(x)=x^4-x^2$ will have two stable fixed points and one unstable fixed point
exhibiting bistability. A well known paradigmatic oscillator with this type of potential is the
classical model of Duffing oscillator~\cite{ikmjb2011}, the ensemble of which with
a common force is represented by 
\begin{align}
	\ddot{x}_i+\alpha\dot{x}_i-\omega_0^2x_i+\beta x_i^3 &- f\sin(\omega t) \nonumber\\ 
	&  =f_e\sin(\omega_e t), i=1,2,\ldots,N 
	\label{duff}
\end{align}
where $\alpha=0.5, \omega_0=1, \beta=1, \omega=1$ and $f=0.33$  are the system parameters
for which the individual Duffing oscillator,  in the absence of the external force ($f_e=0$), 
exhibits periodic oscillations with period $T\approx\frac{2\pi}{\omega}$, 
$\omega_e=0.5$ is the frequency of the external forcing.  We have
fixed the number of oscillators as $N=500$ and the system parameters as above 
throughout the manuscript. The Duffing oscillator has been investigated in great detail 
for its dynamical behavior with and without an
additional external forcing for its chaotic nature and resonance phenomenon~\cite{ikmjb2011}.

Initial conditions of the ensemble of identical oscillators are uniformly  
distributed in both the wells with equal probability in our simulation.
Without any additional external force ($f_e=0$) the ensemble of individual oscillators 
will be oscillating synchronously in either of the wells 
as depicted in Fig.~\ref{fig1}(a).  However, the oscillators in both the wells exhibit 
out-of-phase oscillations with each other while oscillating with the same frequency $\omega=1$
(see insets of Fig.~\ref{fig1}(a)) due to the symmetry property of Eq.~(\ref{duff}).
With the addition of the common external force
to the ensemble, the synchronized oscillators in one of the wells desynchronize
to form self organized clusters with distinct frequencies 
(see insets of Fig.~\ref{fig1}(b)),  while the oscillators
in the other well remain synchronized as shown in Fig.~\ref{fig1}(b) for $f_e=0.088$. 
Increasing the amplitude of the external force to
$f_e=0.135$, the self organized clusters desynchronize completely  (Fig.~\ref{fig1}(c)). 
Thus the splitting of the basin of attraction of an ensemble of identical oscillators, for appropriate 
values of external forcing amplitude, into coherent and incoherent domains
under the influence of a common external forcing leads to the existence of chimera-like states.

Upon increasing $f_e$ further, we find that  the desynchronized oscillators hop back and forth
to the other well where the synchronized group resides, the snap shot of which is
depicted in Fig.~\ref{fig1}(d) for $f_e=0.15$ again representing the chimera-like state. It is
evident from Figs.~\ref{fig1}(c) and~\ref{fig1}(d) that the desynchronized group
remains confined to one of the wells for $f_e=0.135$ and gets distributed in both the wells for $f_e=0.15$,
thereby distinguishing two distinct chimera-like states {\emph C-I} and {\emph C-II}, respectively.  The phase and frequency
of the desynchronized group are clearly distinct from that of the synchronized group
in the former case (see insets of Fig.~\ref{fig1}(c)), whereas they are distributed about
the phase and frequency of the synchronized group in the latter case as depicted in the insets of Figs.~\ref{fig1}(d) and (e).
Some of the mobilized desynchronous oscillators are trapped in the other well
 to form a large cluster of synchronized oscillators for further larger
values of the amplitude of the external forcing. For $f_e=0.16$, increase in the size of
synchronized group is corroborated from the degree 
of phase and frequency distribution as in  Fig.~\ref{fig1}(e). 
Finally, the desynchronized oscillators are all attracted to the other well
to form  a single synchronized cluster for $f_e=0.19$ as illustrated in
Fig.~\ref{fig1}(f).  Thus the ensemble of bistable oscillators loses its bistability to become monostable by switching
the stability of one of the stable fixed points via the chimera-like states as a function of $f_e$.
Existence of analogous chimera states between two populations with both 
inter- and intra-global couplings, as employed by Abrams et al~\cite{dmarm2008},  has been demonstrated in 
discrete chemical oscillators based on the photosensitive Belousov-Zhabotinsky (BZ) reaction
by Tinsley et al~\cite{mrtsn2012} and in mechanical oscillator networks by Martens et al~\cite{eamst2013}.
Interestingly, we have observed similar transitions  even in the states emerging from the basin of attraction of an ensemble 
of identical bistable oscillators driven by a common force.

\begin{figure}[tbp]
	\centering
	\includegraphics[width=1.0\columnwidth]{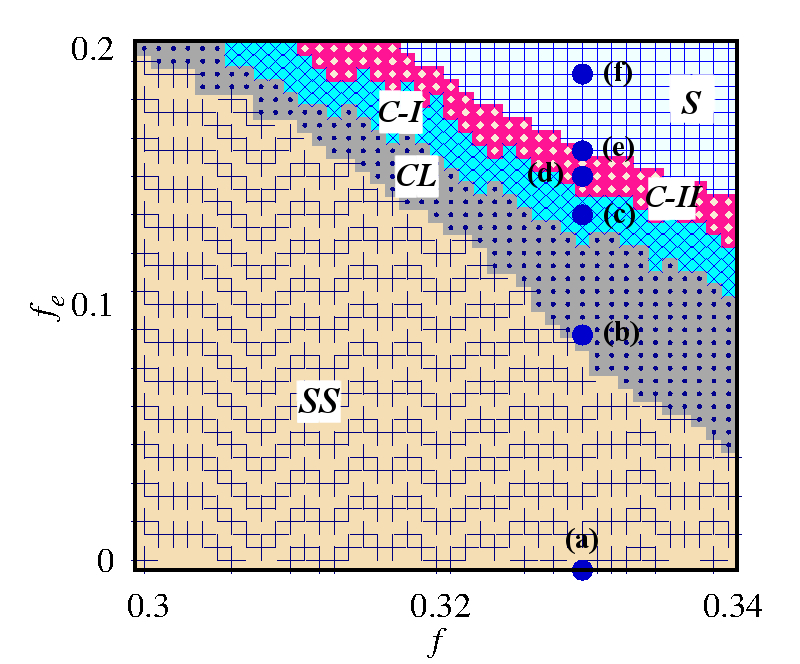}
	\caption{(Color online) Two parameter ($f-f_e$) phase diagram demarcating various dynamical regimes. 
		See text for details.} \label{fig2}
\end{figure}

To obtain a global perspective, we have depicted the $f-f_e$ two parameter phase diagram
in Fig.~\ref{fig2}.
The parameter space corresponding to the collective steady state dynamical behaviors discussed in 
Figs.~\ref{fig1}(a)-(f) are labeled as (a)-(f) in  Fig.~\ref{fig2}.
The steady states exhibit synchronous oscillations in either of the
wells depending on the distribution of the initial conditions in the region marked
as `$SS$'. Clusters of synchronous oscillators are seen in one of the wells in the region 
indicated by $`CL$'. Two distinct chimera-like states are observed in the regions denoted by
`$C-I$' and `$C-II$', respectively. Synchronized single-well oscillations are found in
the  parameter space `$S$'. 

\begin{figure}[tbp]
	\centering
	\includegraphics[width=1.0\columnwidth]{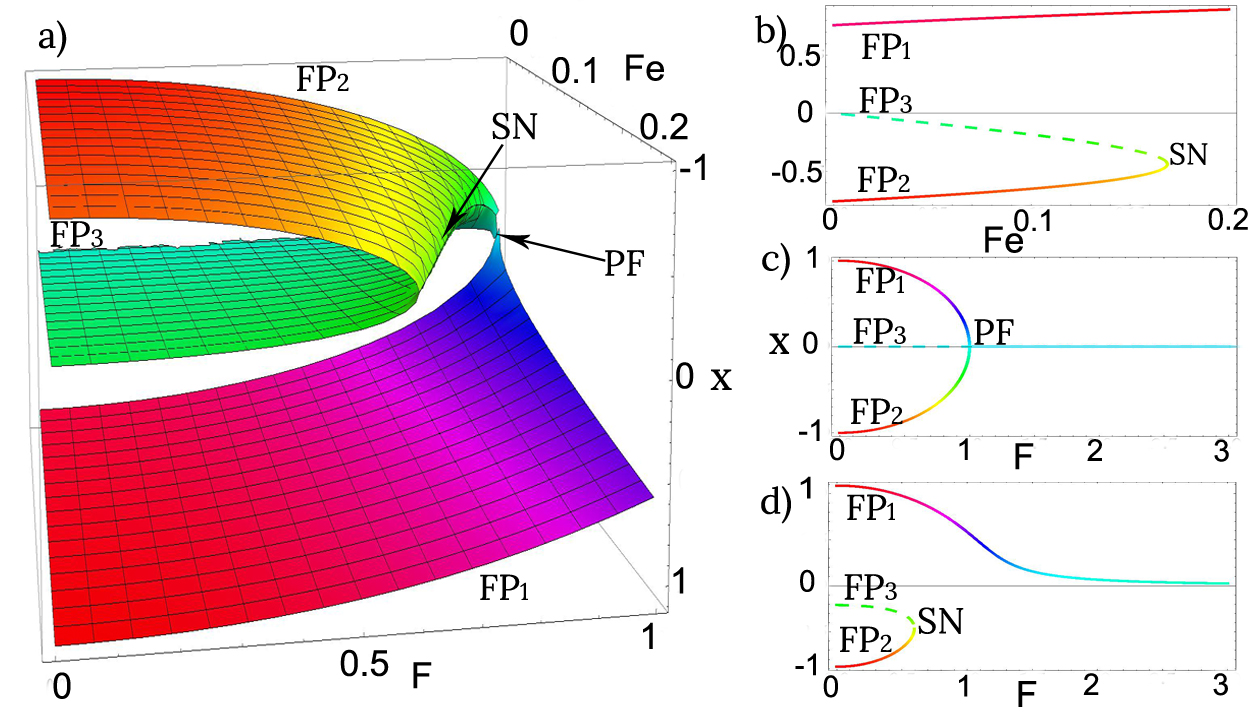}
	\caption{(Color online) Analytical stability diagrams.  (a) Profile of the fixed points of the slow-scale 
		dynamical equation, Eq. (\ref{ampli_eq}), (b) Level curve of Fig.~\ref{fig3}(a) depicting saddle-node bifurcation
		of the ensemble of oscillators for  $F=0.65$,  (c) Level curve of Fig.~\ref{fig3}(a) for $F_e=0.0$
		showing pitch-fork bifurcation of the individual oscillators, and
		(d) Level curve of Fig.~\ref{fig3}(a) for $F_e=0.2$ illustrating  saddle-node bifurcation
		of the ensemble of oscillators. FP1, FP2 and FP3 refer to the
		three fixed point solutions of Eq. (\ref{ampli_eq}).} \label{fig3}
\end{figure}

To explore the reason behind the observed transitions, we start with the 
dynamics of Eq. (\ref{duff}) without any forcing.  For $f=f_e=0$, Eq. (\ref{duff})
has two stable fixed points and one unstable fixed point. Depending upon the distribution of the
initial states, the ensemble of oscillators is attracted towards either of the stable
fixed points.  For finite values of $f$, the   oscillators  corresponding to
the stable fixed points oscillate in opposite phase but with the same frequency, observed
in  Figs.~\ref{fig1}(a) and ~\ref{fig2}(a), in the
asymptotic $(t\rightarrow\infty)$ limit  because of the invariant condition 
$\left(x_i,f\sin(\omega t)\right)\rightarrow\left(-x_i,f\sin(\omega t+\pi)\right)$.
Now, using the multi-time scale perturbation theory~\cite{af2006} and
introducing a small parameter $\varepsilon<<1$ as 
$(\alpha, \omega_0, \beta, f_e)\rightarrow \varepsilon(\alpha, \omega_0, \varepsilon\beta, \varepsilon f_e)$,
the solution
for  Eq. (\ref{duff}) can be separated as $x_i(t)=y_i(\omega t)+z_i(\varepsilon t)$, where 
$y_i(\omega t)=-\frac{f}{\omega^2}\sin(\omega t)$ is the solution of Eq. (\ref{duff}) 
for $f_e=0$ by neglecting the higher harmonics, and $z(\varepsilon t)$ is a slow time scale perturbation.  
Plugging $x_i(t)=y_i(\omega t)+z_i(\varepsilon t)$ in  Eq. (\ref{duff}) and
averaging over $0$ to $\frac{2\pi}{\omega}$, we obtain
\begin{align}
	\ddot{z}_i+\alpha\dot{z}_i+(F^2-\omega_0^2)z_i+\beta z_i^3=F_e,
	\label{ampli_eq}
\end{align}
where $\left<x\right>=\frac{\omega}{2\pi}\int_0^{\frac{2\pi}{\omega}}x dt=z$, $F^2=\frac{3\beta f^2}{2\omega^4}$
and $F_e=\left<f_e\sin(\omega_e t)\right>$ is non-zero for non-integer values of $\frac{\omega_e}{\omega}$
breaking the reflection (left-right) symmetry leading to saddle-node bifurcation around which the
dynamical transitions occur.
Equation~(\ref{ampli_eq}) has three 
fixed points (a saddle and two stable foci/nodes)
for $F_e<F_{c,\pm}=\pm\frac{2(\omega_0^2-F^2)^{\frac{3}{2}}}{3\sqrt{3\beta}}$. 
The stability and the nature of the above fixed points determines the observed dynamical transitions discussed
in Figs.~\ref{fig1} and~\ref{fig2}.  Profiles of the 
fixed points  of Eq.~(\ref{ampli_eq}) is depicted in Fig.~\ref{fig3}(a) as a function of 
$F$ and $F_e$. The stable fixed points are symmetric about the saddle for 
$F_e=0$ (see Fig.~\ref{fig3}(a)), whereas upon increasing $F_e$ one of the stable (negative) fixed points
and the saddle converge towards each other to merge at $F_e=F_{c,+}$ via
saddle-node bifurcation as depicted in Fig.~\ref{fig3}(a). Conversely, the stable (positive)
fixed point and the saddle can also converge towards each other to merge at 
$F_e=F_{c,-}$. This means if  $F$ is replaced by $-F$, then we observe similar dynamics
but in the other well.  The fixed point profile for $F=0.65$ is shown in 
Fig.~\ref{fig3}(b) as a function of $F_e$. It is evident from both Figs.~\ref{fig3}(a) and (b) that the
separation between the saddle and the stable (negative) fixed point decreases monotonically 
upon increasing $F_e$, while the other fixed point remains unaltered. The degree of separation and
the distribution of the initial states of the ensemble of oscillators altogether determine the
nature of dynamical transitions as a function of the forcing. 

Cluster states are formed in a range of $F_e$ for fixed $F$ depending on the 
degree of repulsion exerted  by the saddle on 
the ensemble of oscillators whose initial states are distributed in the vicinity of
the saddle and the stable fixed point. Such cluster states are represented 
in Fig.~\ref{fig1}(b).
Indeed cluster states are formed  in both the wells even for $F_e=0$ ($f_e=0$) but for large values of $F$ $(f)$ as can
be seen in Fig.~\ref{fig2}.  However,  states mimicking chimera do not exist for $F_e=0$  
because both the fixed points approach the saddle resulting in complete incoherent oscillations in
both the wells for large $F$~\cite{ikmjb2011}. Finally, the fixed points collide with the
saddle to create a new stable fixed point through pitch-fork bifurcation (see Fig.~\ref{fig3}(c)) resulting in
double-well oscillations. In contrast, for a finite value of $F_e$, above $F_e=0$, the ensemble of
oscillators undergo double-well oscillations  via cluster, chimera-like states and single-well oscillations,
where the symmetry between the fixed points are broken and  saddle-node bifurcation occurs 
at the onset of single-well oscillations as a function of $F$ as depicted
in Fig.~\ref{fig3}(d) for $F_e=0.2$.

As the separation between the saddle and the stable fixed point narrows down further 
on increasing $F_e$ beyond that of the cluster states, the saddle
expels the oscillators away from the stable fixed point resulting in incoherence
among the oscillators in the well, while the oscillators in the other well are 
in coherence by which {\emph C-I} emerges in Fig.~\ref{fig1}(c). 
For further large values of $F_e$ the separation between the saddle and the stable fixed point
becomes negligibly small and they eventually merge together to form a saddle-node 
bifurcation. In this narrow range of $F_e$ (as seen in 
the two parameter phase diagram in Fig.~\ref{fig2}), some of the oscillators whose initial conditions
lie in the vicinity of the saddle makes round trip among both the wells and some of them are attracted
in their opposite well depending on the value of $F_e$ elucidating the existence of {\emph C-II} states,
which we have observed in Figs~\ref{fig1}(d)-(e).  Beyond the saddle-node bifurcation,
the saddle and stable fixed point disappear only with the existence of the stable positive fixed
point in the other well to which all of the oscillators are attracted finally for sufficiently large
$F_e$ confirming our discussion on Fig.~\ref{fig1}(f).

\begin{figure}[tbp]
	\centering
	\includegraphics[width=1.0\columnwidth]{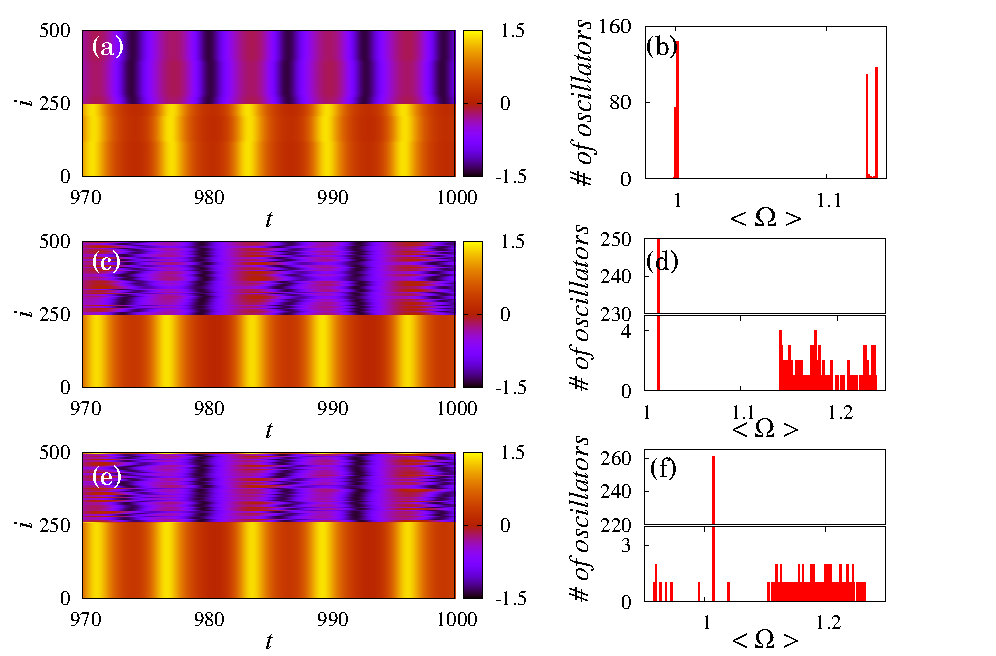}
	\caption{(Color online) Spatiotemporal plots (left column) and  probability 
distribution of time averaged frequency (right column) of an ensemble of Duffing oscillators 
with linear nearest neighbor coupling in the presence of the common external forcing
depicting (a)-(b) cluster states for $\varepsilon=0.06$ and $f_e=0$,
(c)-(d) chimera-like states for $\varepsilon=0.0$ and $f_e=0.14$, and
(e)-(f) chimera states for $\varepsilon=0.06$ and $f_e=0.14$.} \label{fig5}
\end{figure}
Now, we will show that even a linear nearest neighbor coupling can lead to the
manifestation of chimera states in an ensemble of Duffing oscillators in the
presence of the common external forcing represented as
\begin{align}
	\ddot{x}_i+\alpha\dot{x}_i-\omega_0^2x_i+\beta x_i^3 &-f\sin(\omega t) \nonumber\\
& +\varepsilon G=f_e\sin(\omega_e t),  i=1,2,\ldots,N 
	\label{duff_coupled}
\end{align}
where $\varepsilon$ is the coupling strength and $G$ is the coupling function characterizing
the linear nearest neighbor coupling $\left(x_{i+1}+x_{i-1}-2x_i\right)$ while the system 
parameters are the same as in Eq.~(\ref{duff}). Spatiotemporal plots and the
probability distribution of the time averaged frequency $\left<\Omega\right>$ are depicted
in the left and the right columns of Fig.~\ref{fig5} for various values of the system parameters.
In the absence of the common external force $f_e=0$
the ensemble of Duffing oscillators with nearest neighbor coupling displays cluster states as shown in 
Figs.~\ref{fig5}(a) and (b) for the coupling strength $\varepsilon=0.06$. The ensemble will
exhibit chimera-like states, as discussed above, when driven by the common external forcing in the absence of
any couplings ($\varepsilon=0.0$) between the oscillators (see  Figs.~\ref{fig5}(c) and (d)) for $f_e=0.14$.  
Chimera states
manifest even due to a linear nearest neighbor coupling in the presence of the common external forcing
as shown in Figs.~\ref{fig5}(e) and (h).  
The spatiotemporal plot and the probability distribution of $\left<\Omega\right>$  is  depicted in 
Figs.~\ref{fig5}(e) and (f), respectively, for the coupling strength $\varepsilon=0.06$
and the forcing amplitude $f_e=0.14$.
It is clear from these figures that there is a simultaneous emergence of
synchronized and desynchronized  domains among the ensemble of identical Duffing
oscillators elucidating the existence of chimera states due to the nearest neighbor coupling
in the presence of the common external force in the same parameter regime where 
chimera-like states are observed without coupling.  The effect of nearest neighbor
coupling is clearly evident from  Fig.~\ref{fig5}(f), where there is a shift 
in the frequency of the synchronized domain compared to that of the chimera-like 
states in Fig.~\ref{fig5}(d), while
the frequency of the desynchronized domain has a much wider distribution than that of the
chimera-like states.  Thus it is evident that the chimera state 
manifests due to the nearest neighbor coupling but in the presence of the common external force.
Chimera states emerge even for such a low value of coupling strength $\varepsilon=0.06$,
essentially because of the precursor chimera-like states in the absence of coupling.
The above results indicate that even an ensemble of oscillators with linear nearest neighbor
coupling can exhibit chimeras under suitable conditions.
Note that  the system (\ref{duff_coupled}) can also be interpreted as an ensemble of
Duffing oscillators with double forcing and with linear nearest neighbor coupling.
We have further verified the generic nature of our results in 
ensembles of Stuart-Landau oscillators and R\"ossler oscillators in the presence of
the common external forcing, the details of which will be published elsewhere. 

To summarize, surprisingly we have found that the basin of attraction of
an ensemble of identical bistable oscillators can
indeed display coexisting coherent and incoherent domains with distinctly different
nature of attractors mimicking chimera states under a common forcing,
even without any explicit coupling, using Duffing oscillator as a typical example.
The common force facilitates a global saddle-node bifurcation of the
ensemble by breaking the symmetry of the individual oscillators exhibiting
pitch-fork bifurcation in the absence of the external forcing. 
The spontaneous splitting of the ensemble is found to emerge near the 
saddle-node bifurcation  under an appropriate forcing.
We have also provided an appropriate
analytical treatment of the original system of equations, where we have separated
the slow-scale dynamics and investigated the stability of the underlying fixed points,
which provides necessary explanation confirming the observed dynamical transitions 
in the simulation.
Further, it is also shown that even a linear  nearest neighbor coupling can lead to the manifestation of the chimera states
in an ensemble of identical Duffing oscillators but in the presence of the common external forcing. 
As driving/influencing others is a natural behavioral tendency in ecology such as in a herd of sheep, 
flock of birds, colony of ants, hive of bees and many more, as well as in epidemics, in neuroscience,
in social networks, etc., there lies every possibility that
such an emergent behavior may exist in several natural systems.  Our results
may open up potential activities in the identification of chimera states in 
appropriate natural systems.
In particular, our results have elucidated that still there are new avenues open  to
extend the horizon of the existence criteria for the framework of chimera 
with more simple and realistic couplings/situations that fits with the real world 
examples.  More significantly, our results will serve as a basic framework  to ensure
the existence of chimeras in laboratory systems before performing experiments using
ensembles of such systems.

The work of VKC is supported by INSA young scientist project.
DVS is supported by the SERB-DST Fast Track scheme for young
scientist under Grant No. ST/FTP/PS-119/2013.
ML is supported by a Department of Science and Technology (DST), Government
of India, IRHPA research project. ML is also supported by a DAE Raja Ramanna
Fellowship.


\end{document}